\documentclass[a4paper]{jpconf}
\usepackage[utf8]{inputenc}
\usepackage{siunitx}
\usepackage{graphicx}
\usepackage[caption=false]{subfig}
\usepackage[capitalize]{cleveref}
\usepackage{placeins}
\usepackage{lineno}

\begin{document}

\title{High-energy particle physics with IceCube}
\author{Tianlu Yuan for the IceCube collaboration}
\address{Dept.\ of Physics \& Wisconsin IceCube Particle Astrophysics Center, University of Wisconsin, Madison, WI 53706, USA}
\ead{tyuan@icecube.wisc.edu}

\begin{abstract}
While the Standard Model has experienced great predictive success, the neutrino sector still holds opportunities for surprises. Numerous ongoing and planned experiments exist to probe neutrino properties at low energies. The IceCube Neutrino Observatory, comprised of over 5000 photomultiplier tubes (PMTs) situated in a cubic-kilometer of ice at the geographic South Pole, lies in a unique position to measure neutrinos at energies of a TeV and higher. In these proceedings, I discuss several exciting particle physics measurements using IceCube data and probes of physics beyond the Standard Model.
\end{abstract}

\section{Introduction}
\label{sec:intro}

There is rich physics accessible through natural sources of neutrinos. At the highest energies, they are both unique messengers from the centers of galaxies and probes of fundamental physics in and beyond the Standard Model. Since their flux falls rapidly, a large volume detector is needed for measurements at the TeV scale and higher. The IceCube Neutrino Observatory is one such instrument, a cubic-kilometer neutrino detector installed in the ice at the geographic South Pole~\cite{Aartsen:2016nxy}, between depths of \SI{1450}{\m} and \SI{2450}{\m}, completed in 2010. With a sensitivity spanning six orders of magnitude in energy, it has discovered PeV astrophysical neutrinos and measured neutrino oscillation parameters with O(\SI{10}{\giga \eV}) atmospheric neutrinos~\cite{Aartsen:2013jdh,Aartsen:2017nmd}. Here, we focus on its capabilities for particle physics measurements with neutrinos above the TeV-scale.

\section{Neutrino-nucleon cross section}
\label{sec:nuxs}

With the exception of the Glashow resonance, neutrino-nucleon interactions are how we detect neutrinos in IceCube. Interactions almost always occur via Deep Inelastic Scattering (DIS) off partons in the nucleon. Above $E_\nu \sim \SI{100}{\tera \eV}$, they are likely to interact in the Earth before reaching the detector. A modification to the cross section will lead to a distinctive change in the arrival flux of upgoing neutrinos. Under the assumption of the Preliminary Reference Earth Model (PREM)~\cite{Dziewonski:1981xy}, and a single-power-law astrophysical flux, this allowed IceCube to measure the neutrino-nucleon cross section with a sample of upgoing, predominately muon neutrinos with two years of data-taking in the 79 string configuration~\cite{Aartsen:2017kpd}. More recently, two efforts have been made to measure the cross section using different samples, with full-sky information and as a function of energy.

\begin{figure*}[htp]
\centering
    \subfloat{
        \includegraphics[width=0.49\textwidth]{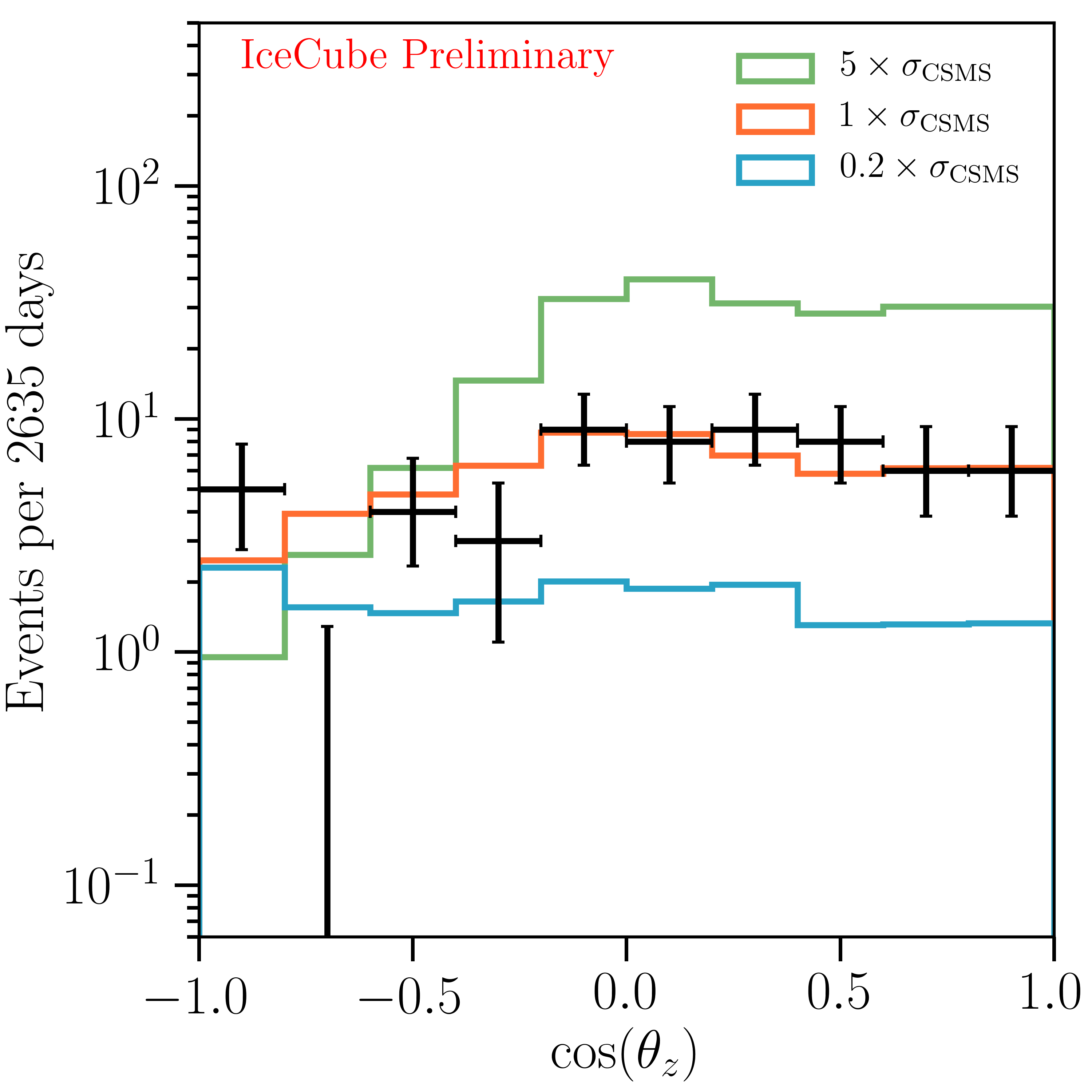}
    }
    \subfloat{
        \includegraphics[width=0.49\textwidth]{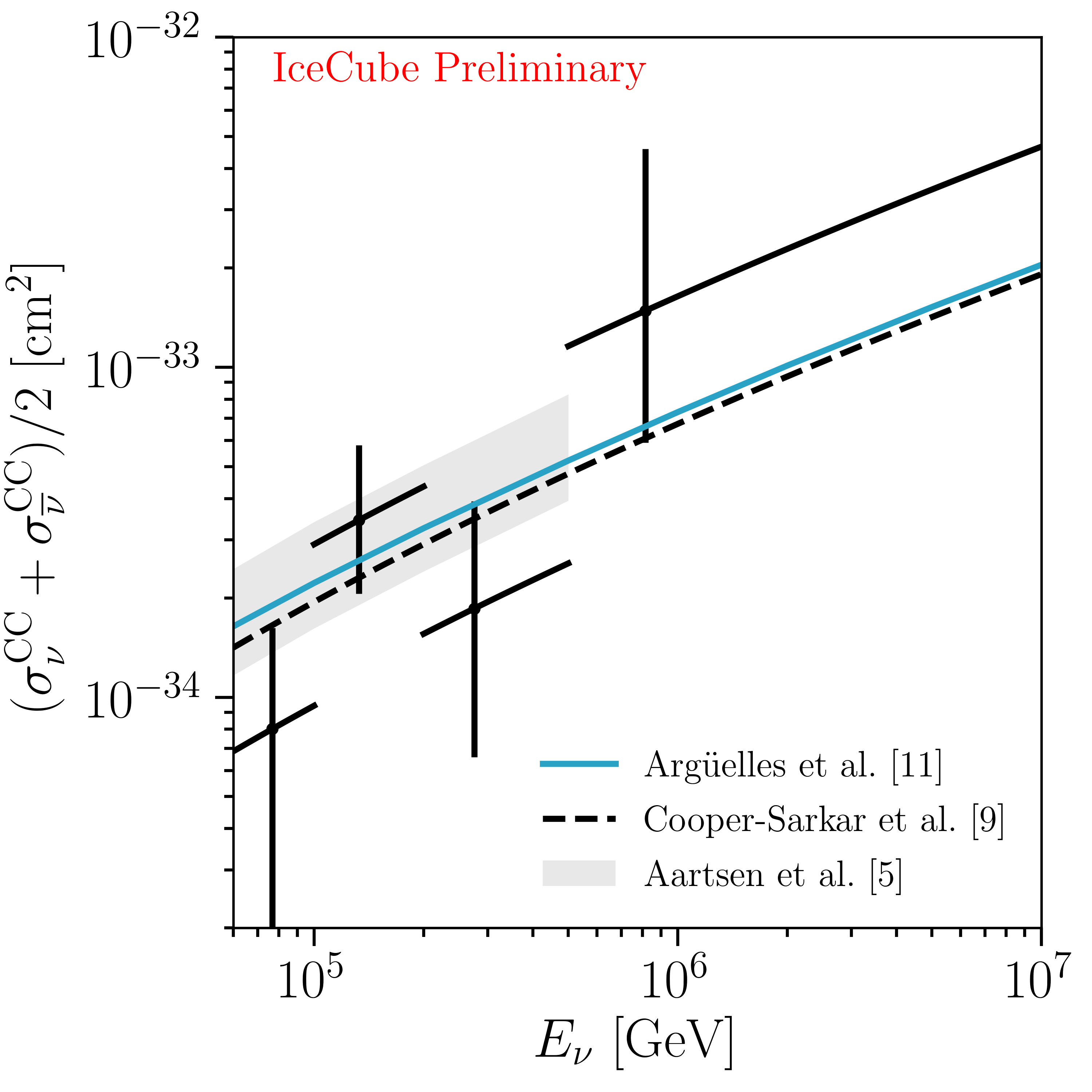}
    }

	\protect\caption{Left panel: Effect of changing the DIS cross section on the expected zenith distribution. Data from the HESE sample are shown as black error bars, while expectations under the nominal DIS cross section are shown in orange. Right panel: The black error bars show 68.3\% confidence intervals, assuming Wilks' theorem, as measured using the HESE sample. The throughgoing $\nu_\mu$ result from~\cite{Aartsen:2017kpd} (shaded gray) is included for comparison. The dashed black line shows predictions from CSMS ~\cite{CooperSarkar:2011pa}. The solid blue line shows $\sigma_\nu^{\mathrm{CC}}$ as calculated in~\cite{Arguelles:2015wba}.}
    \label{fig:hese}
\end{figure*}

One approach relies on the high-energy starting events (HESE) sample with 7.5 years of data-taking~\cite{Schneider:2019ayi}. This veto-based selection allows for full-sky coverage and contains information from particles identified as tracks, cascades or double cascades~\cite{Schneider:2019ayi, Usner:2018qry}. A measurement using only cascades from the 6-year HESE sample was performed independently in~\cite{Bustamante:2017xuy}. Again, we assume a single-power-law astrophysical flux and the PREM density. The ratio of the charged-current (CC) to neutral-current (NC) cross section and the ratio of the neutrino to antineutrino cross section are fixed to their next-to-leading order predictions by Cooper-Sarkar, Mertsch and Sarkar (CSMS)~\cite{CooperSarkar:2011pa}. The cross section is then measured as a function of $E_\nu$, the neutrino energy, in four bins from \SI{60}{\tera \eV} to \SI{10}{\peta \eV} by reweighting events using the nuSQuIDS package~\cite{Delgado:2014kpa} and performing a forward-folding fit of distributions in reconstructed energy, zenith, and topology. Regeneration from NC and tau neutrino interactions are taken into account by nuSQuIDS. The effect of changing the DIS cross section on the zenith distribution is shown in the left panel of \cref{fig:hese} and results, consistent with predictions from the SM, are shown in the right panel of \cref{fig:hese}.

Another approach is to compute the ratio of downgoing vs. upgoing events as a function of the cross section. Using this, one can map the observed ratio to a cross section~\cite{Xu:2018zyc}. This analysis used a sample of contained cascades over four years of data-taking, which typically correspond to either electron or tau neutrino interactions or neutral current interactions from all flavors~\cite{Niederhausen:2017mjk}. Regeneration by $\nu_\tau$ was shown to have a small effect for this analysis and is therefore neglected. For the cross section measurement, an iterative approach is used to unfold reconstructed distributions into distributions in the space of true neutrino energy and direction. This allows us to construct the ratio as a function of $E_\nu$ and thereby measure the cross section in four bins from \SI{6}{\tera \eV} to \SI{10}{\peta \eV}.

\section{Inelasticity}
\label{sec:inel}

IceCube has also measured the inelasticity distribution of neutrino interactions~\cite{Aartsen:2018vez}. The inelasticity for muon neutrino charged-current events, $y$, is defined as the fraction of the neutrino energy transferred to hadrons. By selecting tracks that start in the detector we can reconstruct the visible cascade energy as compared to its total estimated energy. Starting tracks and cascades were selected using a veto-based criteria similar to that used by the HESE selection, but with a lower threshold. This extends the sensitivity of this analysis down to \SI{1}{\tera \eV}. A boosted decision tree is then used to further reject backgrounds from atmospheric muons, while classifying the signal as either cascades or tracks. This leaves a sample of 965 cascades and 2650 tracks.

\begin{figure*}[htp]
\centering
\includegraphics[width=0.5\textwidth]{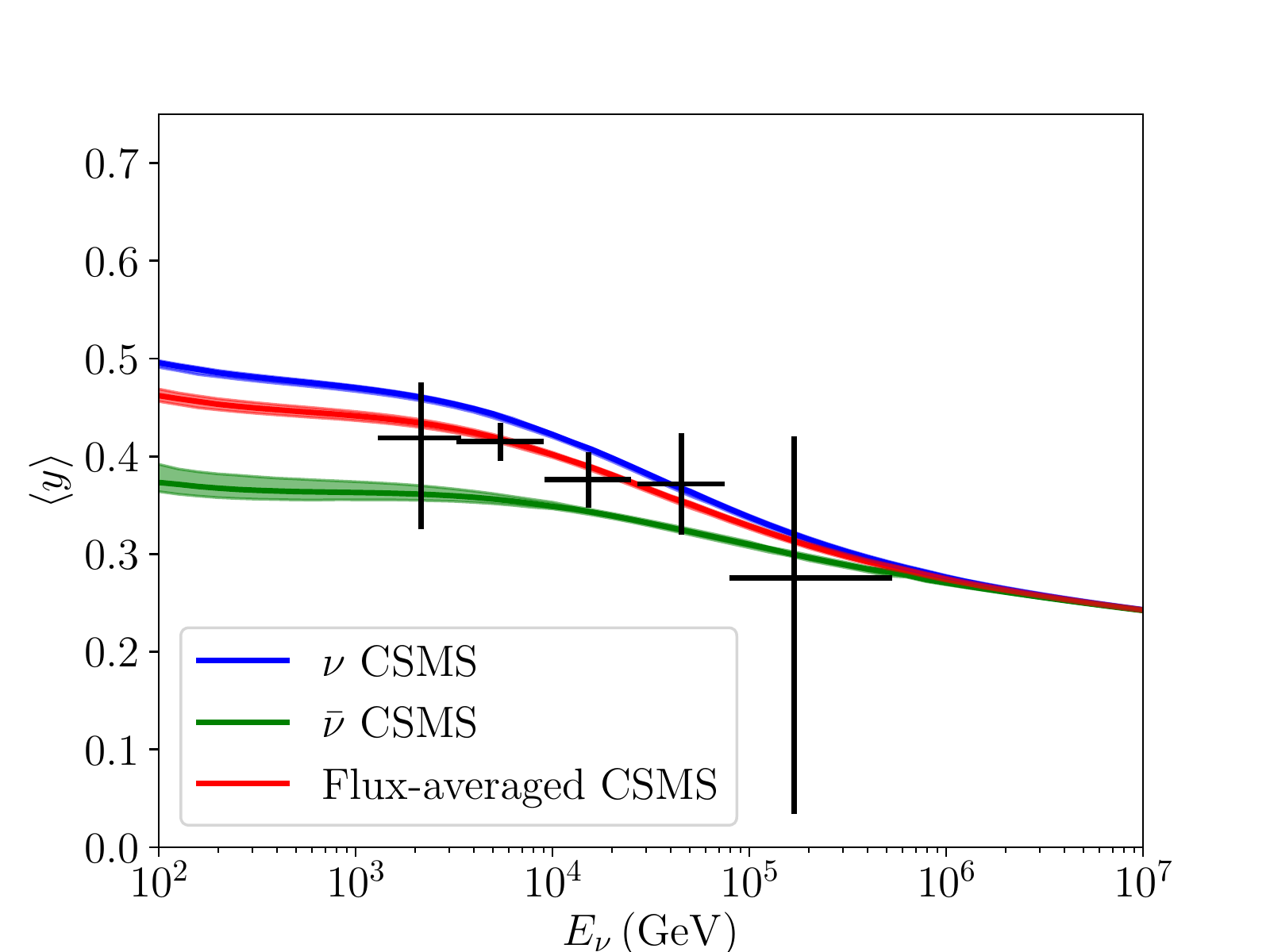}
	\protect\caption{The mean inelasticity as measured and described in~\cite{Aartsen:2018vez}. Predictions from CSMS for neutrinos (blue) and antineutrinos (green) are also shown~\cite{CooperSarkar:2011pa}.}
    \label{fig:inel}
\end{figure*}

A random forest is trained to output the visible cascade energy $E_{\rm casc}$ and the estimated track energy $E_{\rm track}$, allowing the computation of $y_{\rm vis} = E_{\rm casc}/(E_{\rm casc}+E_{\rm track})$. In order to measure the true inelasticity, a parametrization is constructed in terms of the mean inelasticity, $\langle y \rangle$, which allows the $y_{\rm vis}$ distributions to be reweighted for different values of $\langle y \rangle$. The distribution of $y_{\rm vis}$ is then fit in different ranges of the visible energy to obtain the result shown in \cref{fig:inel}.

\section{Beyond the Standard Model}
\label{sec:bsm}

\begin{figure*}[htp]
\centering
\includegraphics[width=0.49\textwidth]{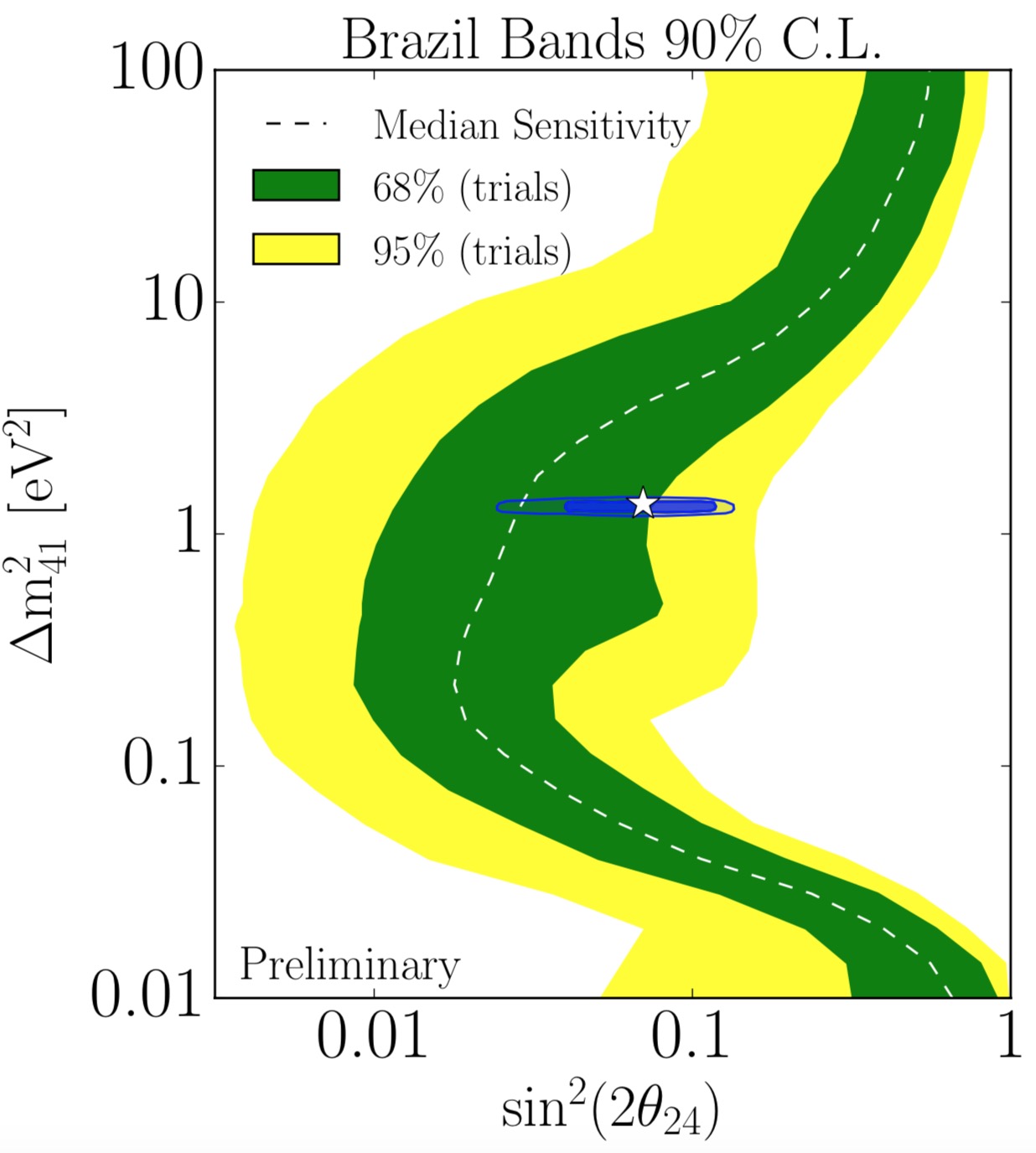}
	\protect\caption{The 68\% (green) and 95\% (yellow) regions of the 90\% confidence limit in the $\Delta m_{41}^2$, $\sin^2 (2 \theta_{24})$ parameter space as evaluated from pseudoexperiments. The blue contour shows the 90\% and 99\% allowed regions from~\cite{Diaz:2019fwt}.}
    \label{fig:sens}
\end{figure*}

Since its completion, IceCube has searched for sterile neutrinos~\cite{TheIceCube:2016oqi}, Lorentz violation~\cite{Aartsen:2017ibm}, magnetic monopoles~\cite{Aartsen:2015exf,Aartsen:2014awd}, non-standard interactions~\cite{Aartsen:2017xtt}, and dark matter~\cite{Aartsen:2018mxl}. Such searches are ongoing and being updated with additional years of data taking. They can set stringent limits on new physics beyond the Standard Model. A recent dark matter search using the HESE 7.5 year sample did not find an abundance of neutrino-dark matter interactions, but set some of the strongest limits on dark matter annihilation and decay~\cite{Arguelles:2019boy}. A update to the sterile neutrino analysis in~\cite{TheIceCube:2016oqi} using 8 years of upgoing muon neutrinos will be able to put stronger bounds on the allowed region of the $3+1$ parameter space. \Cref{fig:sens} illustrates the expected sensitivity in the $\Delta m_{41}^2$, $\sin^2 (2 \theta_{24})$ parameter space.

\section{Conclusion}
\label{sec:fin}

Measurements of the neutrino-nucleon cross section have now been performed using upgoing and all-sky samples, including information from all three flavors. Another approach to measure the cross section using the expected ratio of downgoing vs. upgoing events was also performed, and all results are consistent with predictions from the Standard Model. IceCube has also performed a first measurement of the inelasticity, the fraction of energy transferred to the hadronic shower when an neutrino interacts with a nucleon. In addition, IceCube has placed limits on new physics and continues to improve on existing limits for many topics beyond the Standard Model.

\FloatBarrier

\bibliography{references}
\end{document}